\documentclass[prr,aps,secnumarabic,amsmath,amssymb,twocolumn,notitlepage,longbibliography]{revtex4-2}
\usepackage[colorlinks=true,urlcolor=blue]{hyperref}
\usepackage[normalem]{ulem}
\usepackage[utf8]{inputenc}
\usepackage[english]{babel}
\usepackage[T1]{fontenc}
\usepackage{mathtools}
\usepackage{mathrsfs}
\usepackage{amsfonts}
\usepackage{bbm}
\usepackage{amsthm}
\usepackage{xcolor}
\usepackage{bbold}
\usepackage{bm}
\usepackage{soul}

\definecolor{red}{rgb}{0.75,0,0}
\definecolor{blue}{rgb}{0,0,0.75}
\definecolor{green}{rgb}{0,0.5,0}

\newcommand{\sH}{H_{0}}
\newcommand{\sW}{\Omega_{0}}

\newcommand{\curv}{\kappa}
\newcommand{\Eqref}[1]{Eq.~(\ref{#1})}

\DeclareMathOperator{\area}{Area}

\begin{document}

\title{Polymorphism in tubulin assemblies: a mechanical model}
\author{Ireth Garc\'ia-Aguilar}
\affiliation{Instituut-Lorentz, Universiteit Leiden, P.O. Box 9506, 2300 RA Leiden, The Netherlands}
\author{Steven Zwaan}
\affiliation{Instituut-Lorentz, Universiteit Leiden, P.O. Box 9506, 2300 RA Leiden, The Netherlands}
\author{Luca Giomi}
\email{giomi@lorentz.leidenuniv.nl}
\affiliation{Instituut-Lorentz, Universiteit Leiden, P.O. Box 9506, 2300 RA Leiden, The Netherlands}
\date{\today}

\begin{abstract}
We investigate the mechanical origin of polymorphic structures in two-dimensional tubulin assemblies, of which microtubules are the best known example. These structures feature twisted ribbons, flat tubulin sheets, macrotubules, and hoops, and they spontaneously assemble depending on the chemical environment. Upon modelling tubulin aggregates as minimally anisotropic elastic shells and using a combination of numerical simulations and analytical work, we show that the mechanical strain in tubulin lattices, originating from asymmetries at the single dimer level, naturally gives rise to polymorphic assemblies, among which cylinders and other tubular structures are predominant for a wide range of values of spontaneous curvature. Furthermore, our model suggests that switching the sign of the sheets' spontaneous Gaussian curvature from positive (i.e. sphere-like) to negative (i.e. saddle-like), could provide a possible route to microtubules disassembly.
\end{abstract}

\maketitle

\section{Introduction}

Despite microtubules (MTs) being the most common examples of {\em in vivo} tubulin assemblies, various other structures have been reported in the literature since the discovery of this molecule in the late sixties~\cite{Borisy1967a,Borisy1967b}. Tubulin is a globular protein present in nearly all living cells, where it is most commonly found in the form of dimers of tightly bound $\alpha-$ and $\beta-$tubulin monomers~\cite{Alberts2017}. Each of these monomers can bind Guanosine-5'-1 triphosphate (GTP), but, while this occurs irreversibly at the $\alpha-$monomer, the GTP nucleotide bound to the $\beta-$monomer can be hydrolyzed to Guanosine diphosphate (GDP), thereby giving rise to a two-state system, where each dimer is either non- (GTP-tubulin) or partially (GDP-tubulin) hydrolyzed. GTP-tubulin dimers can then polymerize into polar chains known as {\em protofilaments}, which, in turn, self-assemble in a variety of different conformations by laterally binding to each other. In MTs, this biochemical setup results in a very dynamical structure, where phases of growth (i.e. {\em rescue}) and shrinkage (i.e. {\em catastrophe}) alternate via an intermediate process during which the protofilaments detach from one another and ``peel out'' into ring-shaped oligomers (see {Fig.~\ref{fig:fig1}d-e}). The cycles of catastrophe and rescue are referred to as {\em dynamic instability}, which is the running engine behind the reorganization of MTs in the cell~\cite{Mitchison1984}.

Beside MTs, tubulin is able to from various MT-like as well as other polymorphic assemblies, depending on the chemical environment. 
Already in 1976, Larsson {\em et al}. reported the formation of flat sheets of tubulin in presence of salts of Zn$^{2+}$~\cite{Larsson1976}. In the usual type of buffer and in conditions favoring assembly, when more that $5\times10^{-5}$ moles of Zn$^{2+}$ are present, protofilaments associate laterally into large open sheets~\cite{Gaskin1977} (see {Fig.~\ref{fig:fig1}a}), also in the presence of other cations such as Co$^{2+}$ \cite{Wallin1977}. In contrast to the defined polarity of MTs, protofilaments in these sheets bind alternating the polarity between adjacent protofilaments~\cite{Baker1978}, displaying a more rugged surface due to the asymmetry in lateral association. These sheets can grow to have more than $60$ protofilaments, compared to only $13$ typically found in MTs. In the presence of microtubule-associated proteins (MAPs), some sheets were reported to curve or even close up into macrotubules (see {Fig.~\ref{fig:fig1}b}). Other assemblies with less curvature than MTs are large hoops resulting from tubulin polymerization in presence of glycerol at $25^{\circ}$C and pH $6.6$ \cite{Unger1990}. In these circular structures, as many as $80$ protofilaments assemble in a shallow spiral having sometimes more than $1000$ nm in circumference \cite{Mandelkow1984}. Tubulin rings with a radius of curvature comparable to that of MTs are common residual structures of MT depolymerization so they typically form in destabilizing conditions such as high pH $> 6.0$ or temperatures below $25^{\circ}$C, rolling into double or triple rings in the presence of cations, e.g., with Mg$^{2+}$ at $0^{\circ}$C ( {Fig.~\ref{fig:fig1}e}). Other curved structures include open ribbons which resemble incomplete MT walls, thus C-shaped (see {Fig.~\ref{fig:fig1}f}). Moreover, these can bind together, with varying polarity to form S-shaped ribbons and other curly structures \cite{Karecla1989, Boehm1987}. Assembly in presence of Ca$^{2+}$ and taxol also leads to the formation of exotic structures such as large helical ribbons (see {Fig.~\ref{fig:fig1}g}), but also with curvatures comparable to that of MT \cite{Matsumura1976,Welnhofer1998}, or rolling tighter into more highly curved twisted ribbons at pH $6.0$ and Ca$^{2+}$ concentrations $>10^{-4}$ M ({Fig.~\ref{fig:fig1}h-j}). A survey of these and other anomalous shapes in tubulin aggregates can be found in a review article by Unger {\em et al.}~\cite{Unger1990}.

The problem of the formation of spontaneously curved structures from tubulin assemblies has drawn attention through the years in relation with the previously mentioned dynamical instability~\cite{Wang2005,Gebremichael2008,Igaev2018,Knossow2020,Gudimchuk2021}. As polymerization only occurs in the GTP-bound state, there is generally a cap of GTP-bound tubulin at the tip of an MT, protecting it from disassembly~\cite{Seetapun2012}. However, being polymerization generally slower than GTP hydrolysis, the growing end of a MT becomes eventually rich of GDP, thereby favoring depolymerization. As a result, dynamic instability may be described in terms of the kinetic lag between polymerization and hydrolysis that leads to the presence or lack of the GTP-cap. Yet, cryo-electron microscopy studies suggested that the loss of stability of the tubulin lattice may be facilitated by the inherent spatial curvature of tubulin dimers~\cite{Wang2005,Gebremichael2008}, as schematically represented in~Fig.~\ref{fig:fig1}l. Experimental evidence indicates that, while both GTP and GDP dimers are intrinsically kinked~\cite{Gudimchuk2021,Knossow2020,Igaev2022,McIntosh2018}, GDP-bound tubulin features a larger curvature than GTP-bound tubulin \cite{Ayukawa2021,Nogales2006}, as suggested by the prominent coiling of GDP-rich depolymerizing protofilaments (see {Fig.~\ref{fig:fig1}d}). In addition, hydrolization is believed to increase the stiffness of tubulin dimers, thereby rendering GTP-bound tubulin more flexible~\cite{Igaev2020,Fedorov2019} than GDP-bound tubulin, hence more prone to comply with a wider range of spontaneous curvature~\cite{Ayukawa2021,Igaev2018,Grafmuller2011}. These latter properties, combined with the effect of other environmental cues, could create the conditions for the structural variability observed in tubulin assemblies, both \textit{in vivo} \cite{Chaaban2017,Welnhofer1998,Habura2005} and {\em in vitro} \cite{Unger1990,Matsumura1976,Karecla1989}. 

In this article, we ignore the biochemical origin of spatial curvature of $\alpha\beta$-~dimers to focus on its mechanical contribution to polymorphism in tubulin assemblies. By means of a combination of numerical simulations and analytical work, we show that the mechanical strain in tubulin lattices originating from asymmetries at the single dimer level naturally gives rise to polymorphic assemblies, among which cylinders and other MT-like structures are predominant for a wide range of values of spontaneous curvature. 

The article is organized as follows. In Sec.~\ref{sec:model}, we present our mechanical model for tubulin thin sheets as elastic surfaces, which are subject to structural intrinsic curvature. In Sec.~\ref{sec:developable}, we start with the simpler case of developable sheets, for which analytical calculations provide some insight in the effect of spontaneous curvature terms and mechanical anisotropy. In Sec.~\ref{sec:non_developable}, we use simulations to look at the more interesting case of flexible ribbons which can admit higher in-plane strain in order to better comply to the imposed curvature, resulting in a more nuanced diagram of equilibrium shapes. The simulations for tubulin sheets with a higher degree of stiffness in Sec.~\ref{sec:rigid} provide a numerical parallel to the analytical calculations.  Section~\ref{sec:conclusions} presents concluding remarks.

\section{\label{sec:model}The model}

\begin{figure*}[t]
\includegraphics[width=\textwidth]{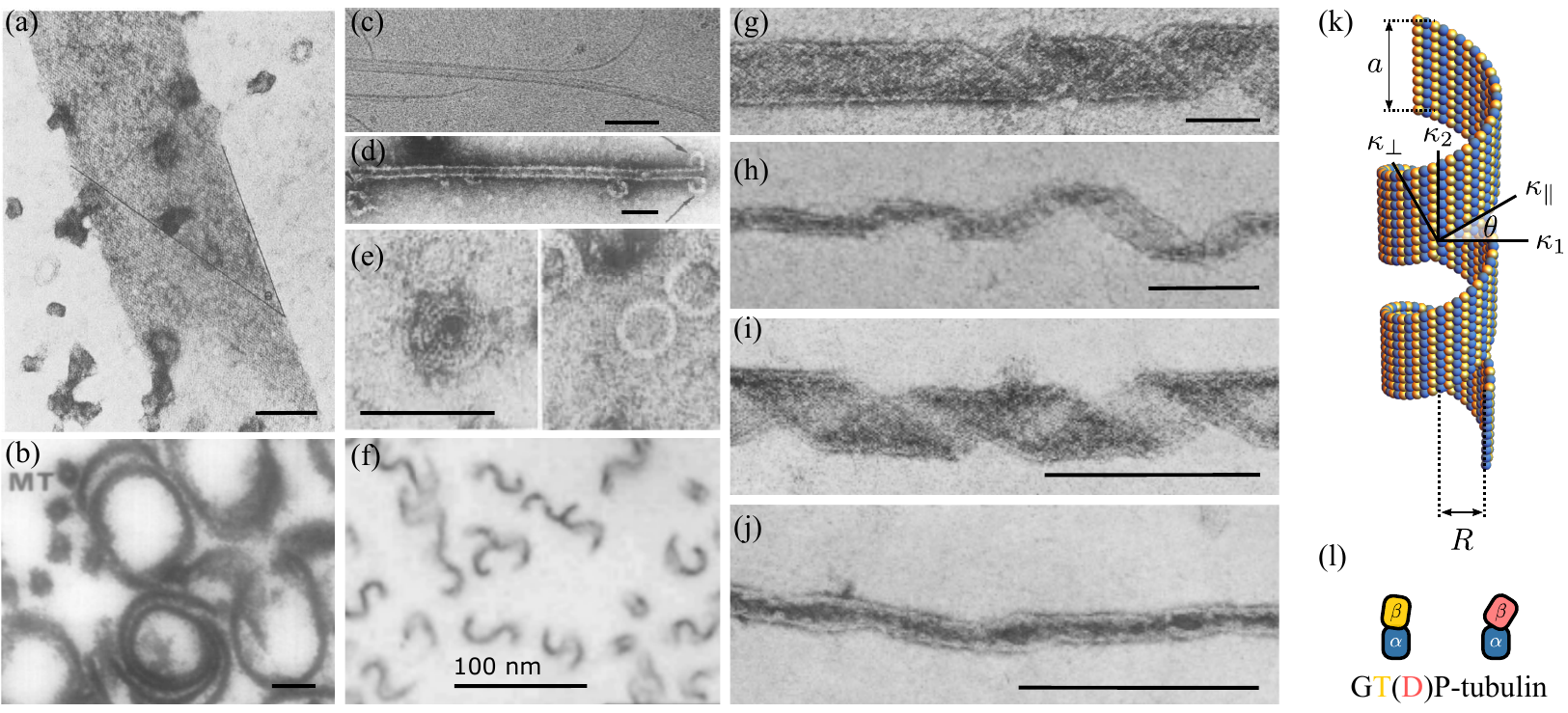}
\caption{\label{fig:fig1} Polymorphism in tubulin structures: (a) Flat sheet, (b) cross-section of macrotubules (see smaller MTs), (c) polymerizing MT with growing curved sheet, (d) depolymerizing MT with peeling ends, (e) ring-shaped oligomers of depolymerized tubulin, (f) C-shaped ribbons bound to each other in different orientations, (g-j) helical ribbons with decreasing radius of curvature, (k) schematic of tubulin assembled in a helical ribbon of radius $R$ and width $a$, (l) GDP-tubulin dimers have a conformational kink giving rise to spontaneously curved protofilaments. GTP tubulin does too, although it is more flexible and likely less curved than GDP-tubulin. All the scale bars in the micrographs correspond to $100$~nm.}	
\end{figure*}

Different mechanical models have been used to study MTs and the sheet structures at their (de)polimerizing ends. For example, cylindrical MTs have been modelled as isotropic thin shells \cite{Sirenko1996,Pablo2003} or spring connected networks \cite{Deriu2007,Schaap2006} to study their elastic properties. The anisotropy induced by asymmetries in dimer conformations was included in work that used orthotropic shallow shells as a model for MT \cite{Wang2006-PhysRevE,Wang2006,Li2006, Qian2007,Huang2008,Yi2008}  or mechanomecanical models of molecular assembly at the tip with tuned interactions \cite{Hunyadi2007, Molodtsov2005-PNAS,Molodtsov2005-BioJ,VanBuren2005,VanBuren2002}. In the continuum elastic model used in Refs.~\cite{Janosi1998,Hunyadi2005} to study the growing ends of MTs, the anisotropy instead is expressed in terms of spontaneous curvatures. Although potentially generic, these models have been restricted to the study MTs and, to the best of our knowledge, none have yet addressed the general polymorphism in tubulin assemblies.

To gain insight into the mechanical origin of polymorphism, we model a tubulin sheet as a two-dimensional crystalline membrane (see e.g. Ref.~\cite{Seung1988}), whose position $\bm{R}=\bm{R}(x^{1},x^{2})$ in the three-dimensional Euclidean space $\mathbb{R}^{3}$ is parametrized in terms of the coordinate $\{x^{1},x^{2}\}$ and whose free energy is given by $F=F_{\rm s}+F_{\rm b}$, where
\begin{subequations}\label{eq:free_energy}
\begin{gather}
F_{\rm s} = \frac{1}{2}\,Y\int {\rm d}A\,\sigma^{2}\;,\\
F_{\rm b} = \int {\rm d}A\,\left[k_{H}(H-H_{0})^{2}+k_{\Omega}(\Omega-\Omega_{0})^{2}\right]\;,
\end{gather}
\end{subequations}
where ${\rm d}A$ is the area of an infinitesimal surface element and the integration spans the entire system. Eq.~(\ref{eq:free_energy}a) is the tubulin lattice stretching energy, with $Y>0$ the Young modulus, $\sigma=g_{ij}\sigma^{ij}/Y$ is the dimensionless trace of the in-plane stress tensor $\sigma_{ij}$ and $g_{ij}=\partial_{i}\bm{R}\cdot\partial_{j}\bm{R}$ the surface metric tensor. The scalar field $\sigma$ is related to the changes in the Gaussian curvature $K=\kappa_{1}\kappa_{2}$, with $\kappa_{1}>\kappa_{2}$ the principal (i.e. maximal and minimal) curvatures, by the Poisson equation (see e.g. Ref.~\cite{GarciaAguilar2020})
\begin{equation}\label{eq:poisson}
\nabla^{2}\sigma = K_{0}-K\;,
\end{equation}
where $K_{0}=c_{1}c_{2}$, with $c_{1}>c_{2}$ are the preferential principal curvatures of the tubulin sheet. 

Equation~(\ref{eq:free_energy}b), on the other hand, is a minimally anisotropic bending energy. Here $H=(\kappa_{1}+\kappa_{2})/2$ is the surface {\em mean curvature} and $\Omega=(\kappa_{1}-\kappa_{2})/2$ the so called deviatoric curvature or {\em warp}. These are in turn related to the Gaussian curvature by
\begin{equation}\label{eq:gaussian_curvature_vs_mean_and_warp}
K = H^{2}-\Omega^{2}\;.
\end{equation}
Notice that, unlike for the Gaussian curvature, the sign of both the mean curvature and the warp is merely conventional and depends upon the orientation of the surface normal vector. Hereafter we will assume both $H_{0}$ and $\Omega_{0}$ positive, so that $c_{1}>0$.

The constants $k_{H}>0$, $k_{\Omega}>0$ quantify the energetic costs associated with a departure from the preferential mean curvature $H_{0}=(c_{1}+c_{2})/2$ and warp $\Omega_{0}=(c_{1}-c_{2})/2$. The latter bending energy was introduced by Fischer in a series of papers in the early nineties~\cite{Fischer1992a,Fischer1992b,Fischer1992c,Fischer1993a,Fischer1993b}, as an alternative to Helfrich's celebrated model of lipid membranes~\cite{Helfrich1973}. In this formulation, evidently inspired by Frank's theory of nematic liquid crystals~\cite{Frank1958}, bending occurs via the superposition of two modes: a purely elliptic one -- associated with the first term in Eq.~{(\ref{eq:free_energy}b)} -- and a purely hyperbolic one -- described by the second term. Each one of these modes has an associated bending stiffness that depends on the material properties. Thus, for a purely elliptic deformation, where $\kappa_{1}=\kappa_{2}$, $F_{\rm b}\sim k_{H}(H-H_{0})^{2}$, whereas for purely hyperbolic, where $\kappa_{1}=-\kappa_{2}$, $F_{\rm b}\sim k_{\Omega}(\Omega-\Omega_{0})^{2}$. 

For specific values of the bending stiffnesses $k_{H}$ and $k_{\Omega}$, and of the curvatures $H_{0}$ and $\Omega_{0}$, the bending energy  Eq.~(\ref{eq:free_energy}b) reduces to that of both lipid membranes and shallow elastic shells. For instance, Helfrich's free energy 
\begin{equation}\label{eq:helfrich}
F_{\rm Helfrich}=\int{\rm d}A\,\left[k(H-c_{0})^{2}+\bar{k}K\right]\;,
\end{equation}
with $k$, $\bar{k}$ and $c_{0}$ material parameters, is readily recovered from Eqs.~\eqref{eq:free_energy} and Eq.~\eqref{eq:gaussian_curvature_vs_mean_and_warp} by setting
\begin{align*}
k_{H} &= k+\bar{k}\;,		& H_{0} &= \frac{k}{k+\bar{k}}\,c_{0}\;, \\
k_{\Omega} &= -\bar{k}\;,	& \Omega_{0} &= 0\;.
\end{align*}
Similarly, the bending energy of a shallow elastic shell of flexural rigidity $D$ and Poisson's ratio $\nu$ is given by (see e.g. Ref.~\cite{Mansfield1989})
\begin{multline}\label{eq:shell}
F_{\rm shell} = \frac{1}{2}\,D\int {\rm d}A
\big[
(\kappa_{\parallel}-c_{\parallel})^{2}+(\kappa_{\perp}-c_{\perp})^{2} \\
+2\nu(\kappa_{\parallel}-c_{\parallel})(\kappa_{\perp}-c_{\perp})
\big]\;,	
\end{multline}
where $\kappa_{\parallel}$ and $\kappa_{\perp}$ are the normal curvatures along two arbitrary orthogonal directions on the mid-surface of the shell (see {Fig.~\ref{fig:fig1}k} and $c_{\parallel}$ and $c_{\perp}$ their corresponding preferential values. Then, calling $\theta$ the angle between, say, the direction associated with the curvature $\kappa_{\parallel}$ and the first principal curvature direction (see Fig.~\ref{fig:fig1}k and using Euler's theorem -- i.e. $\kappa_{\parallel}=\kappa_{1}\cos^{2}\theta+\kappa_{2}\sin^{2}\theta$ -- allows one to express
\[
\kappa_{\parallel} = H+\Omega \cos 2\theta\;,\qquad
\kappa_{\perp} = H-\Omega \cos 2\theta\;,
\]
from which Eq.~\eqref{eq:shell} can be cast in the form given by Eq.~\eqref{eq:free_energy} by setting
\begin{align*}
k_{H} &= D(1+\nu)\;, 						&	H_{0} &= \frac{c_{\parallel}+c_{\perp}}{2}\;,\\
k_{\Omega} &= D(1-\nu)\cos^{2}2\theta\;,	&	\Omega_{0} &= \frac{c_{\parallel}-c_{\perp}}{2}\,\sec 2\theta\;.
\end{align*}
Our model tubulin sheet, whose free energy is given by Eqs.~\eqref{eq:free_energy}, is then equivalent to a shallow elastic shell with a minimal amount of anisotropy built into the preferential curvatures $c_{\parallel}$ and $c_{\perp}$, and the angle $\theta$ expressing the inclination of the protofilaments with respect to the largest principal curvature direction. However, by being expressed in terms the mean curvature $H$ and the warp $\Omega$, it offers the additional advantage of decoupling elliptic and hyperbolic deformations, thus bringing to the forefront the fundamental shape-changing modes of two-dimensional media. Furthermore, while in elastic shells of finite thickness the Poisson ratio is subject to the upper bound $\nu\le 1/2$, resulting from volumetric constraints~\cite{Mansfield1989}, the material parameters $k_{H}$ and $k_{\Omega}$ are amenable to less restrictive interpretations, rooted in the fact that tubulin lattices comprise a single layer or molecular building blocs, to which volumetric constraints do not apply. 

\section{\label{sec:developable}Developable tubulin sheets}

To get a sense of the spectrum of possible shapes accessible to our model tubulin sheet, we start from the simple case of developable surfaces, that is, isometric to a planar rectangle, whose edges have length $a<b$ and whose Gaussian curvature is everywhere vanishing.  Such a class includes, in addition to the flat rectangular sheets, cylinders and the helical ribbons. To this end, we set $K=0$ so Eq.~\eqref{eq:poisson} reduces to
\begin{equation}\label{eq:developable_poisson}
\nabla^{2}\sigma = K_{0}\;.
\end{equation}
Because of developability, all conformations of the tubulin sheet bear the same stress when prevented from adopting their spontaneous Gaussian curvature $K_{0}$. Expressing Eq.~\eqref{eq:developable_poisson} in rectangular coordinates $0\le x \le a$ and $0 \le y \le b$ and solving it with stress-free boundary conditions (see Appendix~\ref{sec:stretching_energy}) gives
\begin{equation}\label{eq:developable_stress}
\sigma = - K_{0}\left(\frac{4}{\pi}\right)^{2}\left.\sum_{n}\right.'\left.\sum_{m}\right.'C_{nm}\sin\left(\frac{\pi n}{a}\,x\right)\sin\left(\frac{\pi m}{a}\,y\right)\;,	
\end{equation}
where the primes indicate that both summations run exclusively on {\em odd} values of the integers $n$ and $m$ and
\begin{equation}
C_{nm}=\frac{1}{nm\left[\left(\frac{\pi n}{a}\right)^{2}+\left(\frac{\pi m}{b}\right)^{2}\right]}\;.	
\end{equation}
Replacing this in Eq.~(\ref{eq:free_energy}a) and computing the integral gives
\begin{equation}\label{eq:developable_stretching_energy}
F_{\rm s} = 32 A \left(\frac{K_{0}}{\pi^{2}}\right)^{2}\left.\sum_{n}\right.'\left.\sum_{m}\right.'C_{nm}^{2}\;,
\end{equation}
where $A=ab$ is the area of the tubulin sheet. In the case of ribbon-like sheets in particular, for which $a\ll b$,  approximating $C_{nm}\approx(a/\pi)^{2}/(mn^{3})$ and summing the series in Eq.~\eqref{eq:developable_stretching_energy} gives
\begin{equation}\label{eq:ribbon_stretching_energy}
F_{\rm s} = \frac{1}{240}\,AY(K_{0}a^{2})^{2}\;.
\end{equation}

By contrast, the bending energy of developable sheets depends on the extrinsic geometry of the system, here embodied by the mean curvature $H$ and the warp $\Omega$. To fix ideas, we consider a generic ribbon-like surface wrapped around a cylinder of radius $R$ (see Fig.~\ref{fig:fig1}e). Developability demands $\kappa_{2}=0$, whereas $\kappa_{1}=1/R$ since, as intuitive, any non-orthogonal planar section of the ribbon has a longer length, thus a smaller radius of curvature, than the orthogonal one. From this we conclude that $H=\Omega=1/(2R)$, so that the bending energy is given by
\begin{equation}\label{eq:developable_bending_energy}
F_{\rm b} = A\left[k_{H}\left(\frac{1}{2R}-H_{0}\right)^{2}+k_{\Omega}\left(\frac{1}{2R}-\Omega_{0}\right)^{2}\right]\;.	
\end{equation}
In the special case of a {\em closed} cylinder, $R=a/(2\pi)$ or $R=b/(2\pi)$, depending on whether the cylinder closes along its longitudinal or transverse side respectively. The bending energy of flat sheets, on the other hand, can simply be recovered by taking the limit $R\to\infty$.

With Eqs.~\eqref{eq:developable_stretching_energy} and \eqref{eq:developable_bending_energy} in hand, one can identify possible equilibrium conformations by minimizing the energy with respect to the radius of curvature $R$ for fixed area $A$ and spontaneous shape curvatures $H_{0}$ and $\Omega_{0}$. As previously mentioned, and as evident from Eq.~\eqref{eq:developable_stretching_energy}, developability guarantees the stretching energy to be shape-independent once the spontaneous Gaussian curvature $K_{0}$ is specified. By contrast, minimizing Eq.~\eqref{eq:developable_bending_energy} with respect to $R$ yields the optimal curvature
\begin{equation}\label{eq:optimal_radius}
\frac{1}{2R} = \frac{k_{H}H_{0}+k_{\Omega}\Omega_{0}}{k_{H}+k_{\Omega}}\;,
\end{equation}
whose corresponding bending energy is given by
\begin{equation}\label{eq:minimal_energy}
F_{\rm b} = \frac{A(H_{0}-\Omega_{0})^{2}}{k_{H}^{-1}+k_{\Omega}^{-1}}\;.	
\end{equation}
Thus, for arbitrary $H_{0}$ and $\Omega_{0}$ values, our model tubulin sheet consists of a developable ribbon, whose mean curvature and warp, which are equal by virtue of the developability constraint, interpolate between $H_{0}$ and $\Omega_{0}$ depending on the magnitude of the corresponding bending stiffnesses. Moreover, if the steric repulsion between tubulin dimers is sufficient to prevent the sheet from forming cigar-like concentric rolls, the maximal principal curvature $\kappa_{1}$ can be at most equal to the curvature $2\pi/a$ of a cylinder of circumference $a$, thus $1/(2R) \le \pi/a$. This upper bound, together with Eq.~\eqref{eq:optimal_radius}, implies that developable tubulin sheets form closed cylinders of radius $2\pi/a$ when
\begin{equation}
\label{eq:W0_inequality}
\Omega_0 \ge \frac{\pi}{a}+\frac{k_H}{k_\Omega}\,\left(\frac{\pi}{a}-H_0\right)\;,
\end{equation}
or even cylinders of smaller radius with overlapping ends when steric repulsion does not prevent rolling. 
For smaller $\Omega_0$ values, helical ribbons instead can roll open or close in order to accommodate the spontaneous curvatures, or else the tubulin sheet bends into a C-shaped ribbon. 
 
The most interesting and possibly counter-intuitive implication of these results is that there exists a regime where a spontaneous negative Gaussian curvature favors cylindrical shapes, namely when both the inequality \Eqref{eq:W0_inequality} and $\Omega_{0}>H_{0}$ hold, since, by virtue of Eq.~\ref{eq:gaussian_curvature_vs_mean_and_warp}, $K_{0}<0$. Although only qualitatively, this feature strengths the idea of protofilaments having a spontaneous negative Gaussian curvature, as inferred by the analysis of the curly shape of the ends of growing MTs~\cite{Chretien1995}. 

\section{\label{sec:non_developable}Non-developable tubulin sheets}

\begin{figure}[t]
\centering
\includegraphics[width=\columnwidth]{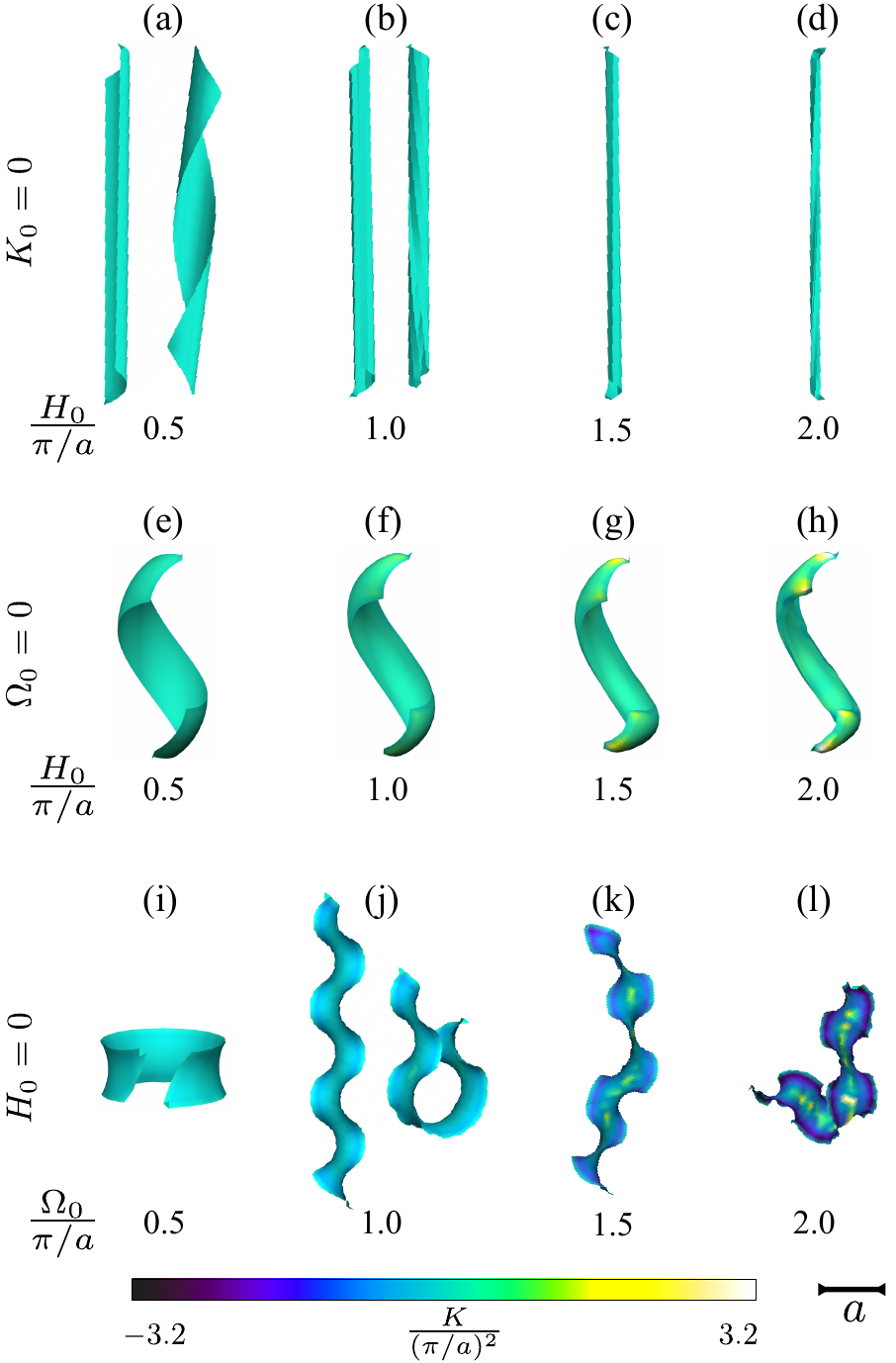}
\caption{\label{fig:3}Equilibrium conformations of an {\em in silico} tubulin sheet obtained from a numerical minimization of the free energy given by Eqs.~\eqref{eq:free_energy}. The three rows correspond respectively to the case of vanishing spontaneous Gaussian curvature (i.e. $K_{0}=0$ and $H_{0}=\Omega_{0}$), vanishing warp (i.e. $\Omega_{0}=0$ and $H_{0}\ne0$) and vanishing mean curvature ($H_{0}=0$ and $\Omega_{0}\ne0$). All configurations shown are to scale, with the width of the initial flat rectangular sheet, $a$, as the scale bar. See Appendix \ref{sec:numerics} for details.}
\end{figure}

In this Section, we lift the constraint of developability and investigate the general case of tubulin sheets with finite Gaussian curvature $K$. To this end we discretize the energies given in Eqs.~\eqref{eq:free_energy} on a triangular mesh and we numerically minimize the resulting multivariate function using gradient descent with adaptive step size (see Appendix \ref{sec:numerics}). For the remainder of this Section, we assume the bending stiffness to be equal and of the same order of magnitude of the energetic cost of stretching: i.e. $k_{H}=k_{\Omega} \sim Ya^{2}$. Furthermore, to achieve an exhaustive sampling of the energy landscape, we initialize our numerical simulations from three different configurations, consisting of a flat rectangular sheet, a helical ribbon and an arch-shaped strip, all having everywhere vanishing Gaussian curvature, thus equal mean curvature and warp. With this setting we investigate four different classes of shapes, for which: {\em 1)} $K_{0}=0$, thus $H_{0}=\Omega_{0}$; {\em 2)} $\Omega_{0}=0$, while $H_{0} \ne 0$; {\em 3)} $H_{0}=0$, while $\Omega_{0} \ne 0$; {\em 4)} the general case where $H_{0}\ne \Omega_{0} \ne 0$. The first three cases are illustrated in Fig.~\ref{fig:3}, whereas the fourth and more general case is summarized in the phase diagram of Fig.~\ref{fig:4}. 

\subsection{\label{sec:zero_gauss}Vanishing spontaneous Gaussian curvature}

When $K_{0}=0$, our model tubulin sheet is energetically favored to relax toward a developable surface, where $H=\Omega$ and the free energy vanishes identically [see Eqs.~\eqref{eq:developable_stretching_energy} and \eqref{eq:minimal_energy}]. Consistently with what we reported in Sec.~\ref{sec:developable}, the lowest energy configuration is degenerate and, for small $H_{0}=\Omega_{0}$ values, can be attained at least by two different configurations, which can be accessed upon initializing the energy minimization from either a flat or twisted initial conformation (Fig.~\ref{fig:3}a and \ref{fig:3}b). For $H_{0}=\Omega_{0}>\pi/a$, however, the degeneracy is broken and our {\em in silico} tubulin sheets roll up in the form of sigar-like multilayered tubes, whose radius decreases monotonically with $H_{0}=\Omega_{0}$ (Fig.~\ref{fig:3}c and \ref{fig:3}d).

\subsection{\label{sec:zero_warp}Vanishing spontaneous warp}

For $\Omega_{0}=0$ and finite $H_{0}$ values, the lowest energy configuration consists, for all initial conformations, of a positively curved helical ribbon, whose principal curvatures $\kappa_{1} \approx \kappa_{2}$ increase with $H_{0}$ (Fig.~\ref{fig:3}e-h). To build up positive Gaussian curvature, the initially flat ribbon stretches (shrinks) along the longitudinal (transverse) direction, while bending in both directions simultaneously. This leads to an overall increase of the mean and Gaussian curvature, while $\Omega$ remains finite, but small in magnitude. A zero free energy configuration, in this case, could be attained upon wrapping the sheet on a sphere of radius $1/H_{0}$, for which $K=K_{0}=H_{0}^{2}$ and $\Omega=\Omega_{0}=0$. Such a configuration, however, is evidently surrounded by local energy minima, which the system can inhabit for an arbitrary long time, thus giving rise to a potentially large spectrum of metastable states. The latter is demonstrated by the residual stretching energy stored in the relaxed configurations.

\subsection{\label{sec:zero_mean}Vanishing spontaneous mean curvature}

When $H_{0}=0$, the sheet relaxes toward a minimal surface (i.e. an area minimizing surface such as a soap-film), having zero mean curvature and negative Gaussian curvature. For small $\Omega_{0}$ values, this is achieved by relaxing towards shapes approximating those of the catenoid (Fig.~\ref{fig:3}c). For large $\Omega_{0}$ values, however, the reduction of the area is no longer compatible with the length of its boundary, which is nearly unstretched, and wrinkles proliferate in the periphery of the sheet (Fig.~\ref{fig:3}j--l), in a way reminiscent of non-Euclidean plates~\cite{Sharon2010}. To clarify this concept, one can use Bernstein-Schmidt isoperimetric inequality on surfaces of constant Gaussian curvature (see e.g. Ref.~\cite{Osserman1978}). That is:
\begin{equation}
\label{eq:isoperimetric}
L^{2} \ge (4\pi-KA)A\;,	
\end{equation} 
where $L$ and $A$ are respectively the perimeter and area of an arbitrary simply connected domain on a surface having constant Gaussian curvature $K$ and the equality holds exclusively in case the domain is by a geodesic disk. In our model tubulin sheet, $L \approx 2(a+b)$ and $K \approx -\Omega_{0}^{2}$. Thus, for large $\Omega_{0}$ values, the inequality in Eq.~\eqref{eq:isoperimetric} cannot longer hold and the sheet buckles into a wrinkled structure, where geometric compatibility is restored because of the larger area covered by the wrinkles.

\subsection{General case}

\begin{figure}[t]
\centering
\includegraphics[width=\columnwidth]{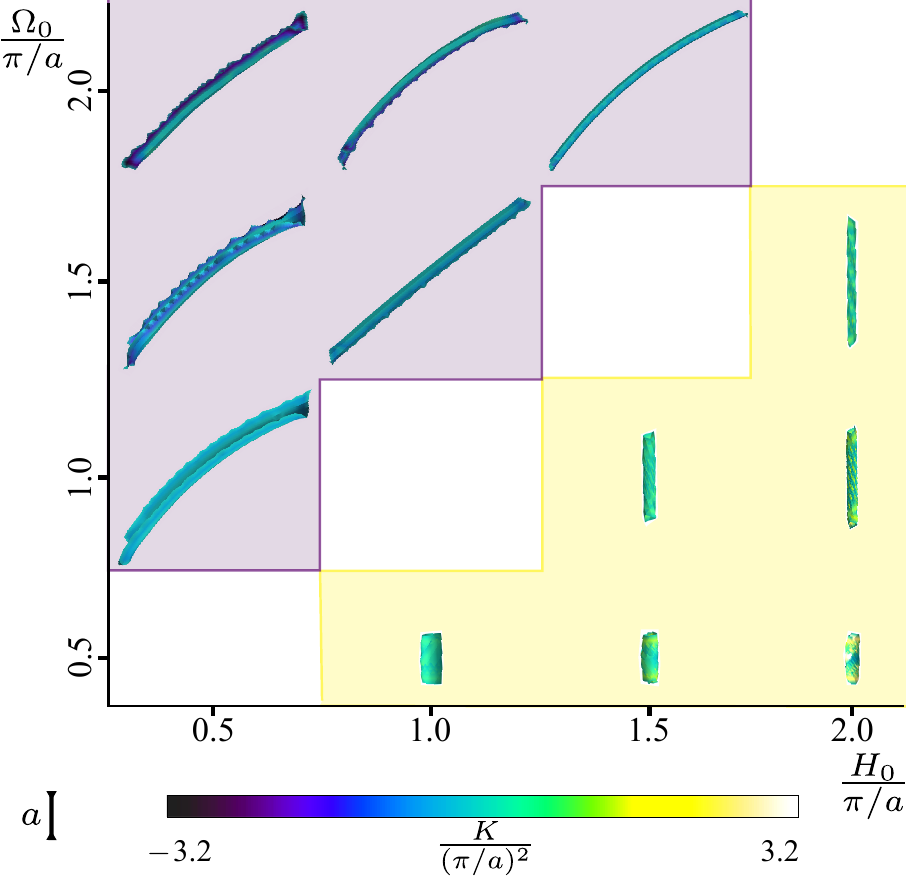}
\caption{\label{fig:4}Phase diagram for sheets with non-vanishing spontaneous Gaussian curvature (i.e. $K_0\neq0$) obtained from a numerical minimization of the free energy given by Eqs.~\eqref{eq:free_energy}. See Appendix~\ref{sec:numerics} for details. The bisectrix, separating regions of positive (i.e. $H_{0}>\Omega_{0}$) and negative (i.e. $H_{0}<\Omega_{0}$) Gaussian curvature, marks a prominent boundary between long cylindrical and multilayered tubular conformations of our model tubulin sheet. All configurations shown are to scale, with the width of the initial flat rectangular sheet, $a$, as the scale bar.}
\end{figure}

Fig.~\ref{fig:4} shows the phase diagram obtained in the general case where both $H_{0}$ and $\Omega_{0}$ are finite, but not equal to each other, so that $K_{0}\ne 0$. Compared to the previous scenarios, here the system is highly frustrated as the non-vanishing spontaneous curvatures render a zero free energy configuration inaccessible to any embedded surface in $\mathbb{R}^{3}$. In order for the free energy to vanish identically, the sheet must indeed have both constant mean and Gaussian curvature, but the only embedded surfaces with this property are patches of either the plane, the cylinder or the sphere. Because in both the plane and the cylinder $K=0$, these shapes correspond to zero free energy configurations respectively for $H_{0}=\Omega_{0}$ (see Sec.~\ref{sec:zero_gauss}) and $\Omega_{0}=0$ (see Sec.~\ref{sec:zero_warp}). For small finite warp and $H_{0}=0$ (see Sec.~\ref{sec:zero_mean}), the lack of zero free energy configurations is compensated instead by the existence of low free energy soap-film-like configurations with vanishing mean and negative Gaussian curvature. The latter are however unavailable when both $H_{0}$ and $\Omega_{0}$ are finite. In particular, when $H_{0}>\Omega_{0}$ -- and a spherical geometry is energetically favored -- the finite spontaneous warp prevents its occurrence. Conversely, for $H_{0}<\Omega_{0}$, the finite spontaneous mean curvature hinders the emergence of a perfect saddles. The minimal free energy configurations found in our numerical simulations consist, in this case, of tightly rolled multilayered cylinders, for $H_{0}>\Omega_{0}$, and shoehorn-shaped saddles, for $H_{0}<\Omega_{0}$. In the latter case, the lack of geometric compatibility between the bulk and the boundary, results in the formation of wrinkles such as those described in Sec.~\ref{sec:zero_mean} for $H_{0}=0$. 

More importantly, our numerical simulations demonstrate how crossing the $H_{0}=\Omega_{0}$ line, from the region where $H_{0}>\Omega_{0}$, drives an unfolding of the tubulin sheet reminiscent of {\em catastrophe} events in MTs and to intermediate sheet structures during assembly \cite{Gudimchuk2021,Nogales2006}. The latter condition can be in principle achieved in different ways at the level of individual protofilaments. Using the equivalence relations derived at the end of Sec.~\ref{sec:model} to express $H_{0}$ and $\Omega_{0}$ in terms of $c_{\parallel}$, $c_{\perp}$ and $\theta$, the inequality $H_{0}>\Omega_{0}$ can be rearranged in the form
\begin{equation}\label{eq:stability}
\cos 2\theta > \frac{c_{\parallel}-c_{\perp}}{c_{\parallel}+c_{\perp}}\;.
\end{equation}
If the spontaneous curvature vanishes along the protofilaments direction (i.e. $c_{\parallel}=0$), Eq.~\eqref{eq:stability} holds for arbitrary $c_{\perp}$ and $\theta$ values. Thus, as long as protofilaments are {\em spontaneously} straight~\footnote{This requires $c_{\parallel}=0$, but $\kappa_{\parallel}\ne 0$ in general.}, helical and tubular configurations are mechanically stable. By contrast, Eq.~\eqref{eq:stability} has no real solutions when $c_{\parallel}$ is finite and of opposite in sign with respect to $c_{\perp}$, since then $K_0<0$. 

In summary, our results suggest that {\em catastrophe} events in MTs could arise from a mechanical instability triggered by a switch of the spontaneous Gaussian curvature $K_{0}$ from positive to negative. Such a switch, in turn, could originate from a conformational change of the tubulin dimers, whose effect is to introduce an arbitrarily small, but finite, spontaneous longitudinal curvature driving the ``peeling'' of the protofilaments away from the MT axis (i.e. $c_{\parallel}<0$ and $c_{\perp}>0$ with the sign convention of this article). This mechanism is consistent with current experimental observations of depolymerized MTs (see Fig.~\ref{fig:fig1}c-d), as well as with the general view on MTs disassembly, which ascribes the occurrence of {\em catastrophe} to a conformational switch from a flexible, lattice-stabilized GTP state, to a strongly radially curved and rigid GDP state~\cite{Stewman2020}.

\section{\label{sec:rigid} Stiff tubulin sheets}

We additionally performed simulations for stiffer tubulin sheets with a Young's modulus 100 times larger than for the non-developable sheets discussed in Sec.~\ref{sec:non_developable}. In this case, the higher cost of stretching limits the transitions between different local minima of the elastic free energy, so that our rigid {\em in silico} tubulin sheets relax almost isometrically. As the initial configurations have $K=0$ by construction (see Appendix~\ref{sec:numerics}), all relaxed configurations are nearly developable, thus acting as a practical validation of the numerical calculations and a handy parallel to contextualize the analytical results presented in Sec.~\ref{sec:developable}.

\begin{figure}[t]
\centering
\includegraphics[width=\columnwidth]{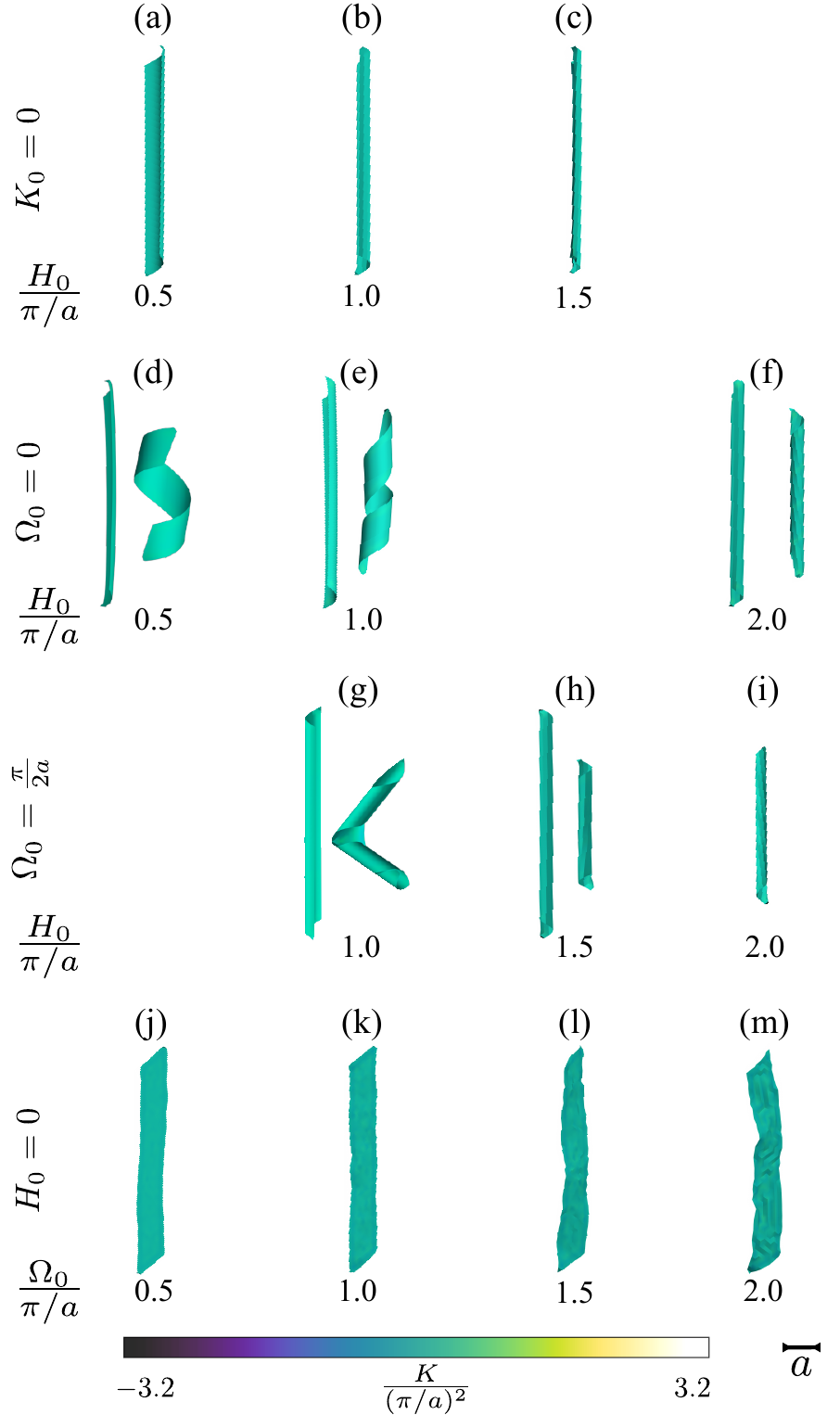}
\caption{\label{fig:rigid}Resulting conformations for simulations of stiff tubulin sheets, where $K \approx K_{0}=0$. The equilibrium conformation have been obtained from a numerical minimization of the free energy given by Eqs.~\eqref{eq:free_energy}, with a Young modulus value 100 times larger than that used in Figs.~\ref{fig:3} and \ref{fig:4}. All configurations shown are to scale, with the width of the initial flat rectangular sheet, $a$, as the scale bar.}	
\end{figure}

Consistently with Eq.~\eqref{eq:W0_inequality}, for $\Omega_0/(\pi/a) \le 2-H_0/(\pi/a)$, we find both helical and C-shaped sheets to be stable, zero-energy configurations, whereas for higher values, only (multilayered) tubular rolls are observed. Still, such multilayered structures have degenerate helical pitch (see Fig.~\ref{fig:rigid}f--h). We note that, despite their Young modulus being two orders of magnitude larger than that used in Sec.~\ref{sec:non_developable}, our model {\em in silico} tubulin sheets are not entirely inextensible. In fact, for the cylindrical and helical shapes in Fig.~\ref{fig:rigid}d--f, the small amount of stretching allowed is expressed as a slight increase of the smallest principal curvature $\kappa_2$, in particular for low $H_0$ values. Because the transverse curvature $\kappa_1$ is proportional to $H_{0}$, the longitudinal curvature $\curv_2 \sim 1/H_{0}$ increases for decreasing $H_{0}$ values, provided that $K=\kappa_{1}\kappa_{2}$ remains small.

When $H_{0}=0$, on the other hand, a large spontaneous warp is accommodated via multiple buckling events, thereby giving rise to networks of wrinkles similar to those found in thin elastic sheets under stress~\cite{Genzer2006}. Lastly, for the general case, we find that, because $K\approx 0$, the rigid sheets fail to comply to the imposed curvature outside of the ${\sH=\sW}$ line, also notable in Eq.~\eqref{eq:optimal_radius}. This explains why rigid sheets with $K_0 \ne 0$ undoubtedly store bending energy. 

\section{Conclusions}

\label{sec:conclusions}

In this article, we explored the occurrence of polymorphism in tubulin assemblies, among which microtubules (MTs) represent the most common and biologically relevant realization. Tubulin sheets are modelled as shallow elastic shells, where {\em unequal} longitudinal and transverse spontaneous curvatures reflect, in a minimal way, the mechanical anisotropy arising from the organization of the $\alpha\beta$ dimers into protofilaments (see Fig.~\ref{fig:fig1}l). 

Unlike previous mechanical models of tubulin sheets (see e.g. Ref.~\cite{Janosi1998, Hunyadi2005}), bending elasticity here is  characterized in terms of mean and deviatoric curvature (or warp), with the goal of bringing to the forefront the fundamental deformation modes of two-dimensional media and highlighting the lack of geometrical compatibility inherent to specific choices of the spontaneous curvatures. As the latter arises at the length scale of the dimers, with the former being a global property of the system, our approach allows us to identify possible regimes where the kinked conformation of tubulin dimers gives rise to prominent geometrical frustration, of which the existence of a rough free energy landscape  -- hence polymorphism -- is the most natural consequence. 

Using combined analytical and numerical work, we showed in particular, how a cylindrical geometry is by far the most robust, for a wide range of spontaneous curvatures. The origin of this robustness is two-fold. On the one hand, in the likely scenario where spontaneous mean curvature and warp are comparable in magnitude (i.e. $H_{0}\approx\Omega_{0}$, but without restrictions about the magnitude or the difference of the spontaneous curvature along the longitudinal and transverse direction of the individual protofilaments), the existence of an intrinsically flat zero free energy configuration (i.e. $K \approx 0$) renders cylinders energetically favorable for a wide range of spontaneous mean curvature and warp values. On the other hand, for $H_{0}>\Omega_{0}$, the finite spontaneous warp renders a zero free energy with constant mean and Gaussian curvature dynamically inaccessible and tubulin sheets are again favored to form (possibly multilayered) tubular structures. 

Finally, we demonstrated that transitioning from $H_{0}>\Omega_{0}$ to $H_{0}<\Omega_{0}$ provides a possible strategy to render the closed tubular conformation unstable to an open one, consistently with experimental observations in MTs after a {\em catastrophe} event.

\acknowledgements

I. G. A. is grateful to Nikita Gudimchuk (Moscow State University), Piermarco Fonda and Ludwig Hoffmannn for discussions and thank Yura Malitsky (Link\"oping University) for the introduction to the adaptive gradient descent without descent and the helpful discussions on its implementation, together with Yevheniia Cheipesh. This work is supported by the Netherlands Organization for Scientific Research (NWO/OCW), as part of the D-ITP program (I. G. A. and L. G.), the Vidi scheme (L. G.), the Frontiers of Nanoscience program (L. G.). 

\appendix

\section{\label{sec:stretching_energy}Stretching energy of developable sheets}

In order to calculate the stretching energy of a developable rectangular sheet, Eq.~\eqref{eq:developable_stretching_energy}, one must first calculate the dimensionless trace $\sigma$ of the covariant stress tensor, which is in turn proportional to the Airy stress function (see e.g. Ref.~\cite{Seung1988}). The latter can be done by integrating Eq.~\eqref{eq:poisson} with Dirichlet boundary conditions: i.e. 
\begin{equation}
\sigma(0,y) = \sigma(a,y) = \sigma(x,0) = \sigma(x,b) = 0\;,
\end{equation}
where $0\le x \le a$ and $0\le y \le b$ are Cartesian coordinates along orthogonal directions parallel to the short and long edges of the sheet, whose length is given by $a$ and $b$ respectively. The integration can be conveniently performed using the Laplacian Green function
\begin{widetext}
\begin{equation}\label{eq:green_function}
G(\bm{r},\bm{r}') = -\frac{4}{ab}\sum_{n=1}^{\infty}\sum_{m=1}^{\infty}\frac{\sin\left(\frac{\pi n}{a}\,x\right)\sin(\frac{\pi n}{a}\,x')\sin(\frac{\pi m}{b}\,y)\sin\left(\frac{\pi m}{b}\,y'\right)}{\left(\frac{\pi n}{a}\right)^{2}+\left(\frac{\pi m}{b}\right)^{2}}\;,
\end{equation}
\end{widetext}
where $\bm{r}=\{x,y\}$ and $\bm{r}'=\{x',y'\}$ are generic points on the sheet. For a generic Gaussian curvature difference $K_{0}-K$ this gives
\begin{equation}\label{eq:poisson_solution}
\sigma(\bm{r}) = \int {\rm d}A'\,G(\bm{r},\bm{r}')\left[K_{0}-K(\bm{r}')\right]\;.	
\end{equation}
Now, in the case of developable sheets, $K=0$ and the integration over the primed coordinates has the effect of removing all terms associated with even values of the integers $n$ and $m$, since
\begin{equation}\label{eq:stress_integrals}
\int_{0}^{a}{\rm d}x'\,\sin\left(\frac{\pi n}{a}\,x'\right)	
= \left\{
\begin{array}{ll}
\frac{2a}{\pi n}\;, & n\,{\rm odd} \\[5pt]
0 					& n\,{\rm even} 
\end{array}
\right.\;.
\end{equation}
Together, Eqs.~\eqref{eq:green_function}, Eqs.~\eqref{eq:poisson_solution} and Eqs.~\eqref{eq:stress_integrals} readily give Eq.~\eqref{eq:developable_stress}. Finally, integrating the square of the stress field and using again Eq.~\eqref{eq:stress_integrals} allows one to obtain the stretching energy given in Eq.~\eqref{eq:developable_stretching_energy}. 

To derive Eq.~\eqref{eq:ribbon_stretching_energy}, on the other hand, one can approximate $C_{nm}\approx (a/\pi)^{2}/(mn^{3})$ under the assumption that $a\ll b$. Then
\begin{gather}\label{eq:stretching_energy_sum}
\left.\sum_{n}\right.'\left.\sum_{m}\right.'C_{nm}^{2} 
\approx \left(\frac{a}{\pi}\right)^{4}\left.\sum_{n}\right.'\frac{1}{n^{6}}\left.\sum_{m}\right.'\frac{1}{m}^{2} \notag \\
= \left(\frac{a}{\pi}\right)^{4}\left[\zeta(6)-\frac{\zeta(6)}{64}\right]\left[\zeta(2)-\frac{\zeta(2)}{4}\right]\notag \\
= \frac{1}{3}\left(\frac{\pi a}{4}\right)^{4}\;,
\end{gather}
where $\zeta(s)=\sum_{n=1}^{\infty}1/n^{s}$ is the Riemann zeta function. Replacing Eq.~\eqref{eq:stretching_energy_sum} into Eq.~\eqref{eq:developable_stretching_energy} finally gives Eq.~\eqref{eq:ribbon_stretching_energy}.

\section{\label{sec:numerics}Numerical simulations}

\begin{figure}[t!]
\centering
\includegraphics[width=\columnwidth]{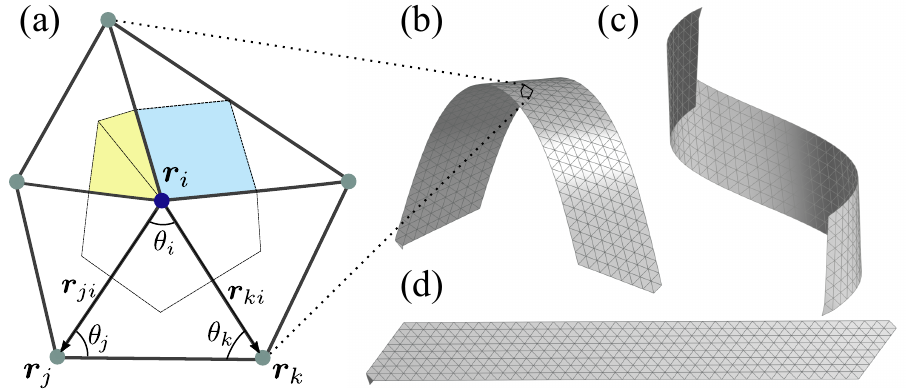}
\caption{\label{fig:5} (a) The discrete fields in Eqs.~\ref{eqs:discrete_energy} for the $i-$th vertex at position $\bm{r}_i$ are expressed as sums over the adjacent triangles, where each triangle $\triangle{ijk}$ is defined for example by the vertices $\{i,j,k\}$. The area element $A_i$ is enclosed with the dashed lines. (c-d) Initial configurations used in the simulations. }
\end{figure}

The equilibrium configurations of our non-developable model tubulin sheets are obtained upon minimizing a discrete variant of the elastic free energy given in Eqs.~\eqref{eq:free_energy} on triangular meshes consisting of $N=369$ vertices of position $\bm{r}_{i}$, $i=1,
\,2\ldots\,N$, and whose topology is fixed during simulations. The discretized free energies are given by
\begin{subequations}
\label{eqs:discrete_energy}
\begin{gather}
F_{s} = \frac{1}{2}k_{S}\,\sum_{\langle ij\rangle}\left(|\bm{r}_{i}-\bm{r}_{j}|-\ell_{0}\right)^{2}\;,\\
F_{b} = \sum_{i}A_{i}\left[k_{H}(H_{i}-H_{0})^{2}+k_{\Omega}(\Omega_{i}-\Omega_{0})^{2}\right]\;,
\end{gather}
\end{subequations}
where $k_{S}$ is a spring constant, proportional to the sheet Young modulus~\cite{Seung1988}, $\ell_{0}$ a rest length setting the overall size of the sheet and $A_{i}$ is the area effectively covered by each vertex. The latter is calculated from the areas of all triangles meeting at the $i-$th vertex using a mixed Voronoi scheme~\cite{Meyer2003}, so that, calling $\triangle{ijk}$ the triangle defined by the vertices $\{i,j,k\}$ whose positions are $\{\bm{r}_{i},\bm{r}_{j},\bm{r}_{k}\}$, one has 
\[
A_{i} = 
\sum_{\triangle{ijk}}
\left\{
\begin{array}{lll}
\frac{1}{2}\,\area(\triangle{ijk}) & & \theta_{i}>\frac{\pi}{2}\\[5pt]
\frac{1}{4}\,\area(\triangle{ijk}) & & \theta_{j},\theta_{k}>\frac{\pi}{2}\\[5pt]
\frac{1}{2}\left(r_{ij}^{2}\cot\theta_{k}+r_{ik}^{2}\cot\theta_{j}\right) & & {\rm otherwise}
\end{array}
\right.,
\]
where $r_{ij}=|\bm{r}_{ij}|=|\bm{r}_{i}-\bm{r}_{j}|$, $\area(\triangle{ijk})=(1/2)\,r_{ij}r_{jk}\sin\theta_{k}$ is the area of $\Delta_{ijk}$ and $\{\theta_{i},\theta_{j},\theta_{k}\}$ are the angles subtended by the three vertices (Fig.~\ref{fig:5}a).

Analogously, the discrete mean curvature $H_{v}$, can be define as
\begin{equation}
H_{i}\bm{n}_{i} 
= \frac{1}{4A_{i}}\,\sum_{\Delta_{ijk}}\left(\frac{\bm{r}_{ij}}{r_{ij}}\,\cot\theta_{k}+\frac{\bm{r}_{ik}}{r_{ik}}\,\cot\theta_{j}\right)\;,	
\end{equation}
where $\bm{n}_{i}$ is the outward-directed unit normal at the $i-$th vertex. The Gaussian curvature, is routinely computed from the deficit angle at each vertex. That is
\begin{equation}
K_{i} = \frac{1}{A_{i}}\,\left[2\pi-\sum_{\Delta_{ijk}} \theta_{i}\right]\;,	
\end{equation}
where the summation runs over all triangles containing the $i-$th vertex. Finally, the discrete warp $\Omega_{i}$ can be calculated, using Eq.~\eqref{eq:gaussian_curvature_vs_mean_and_warp}, from $H_{i}$ and $K_{i}$. That is
\begin{equation}
\Omega_{i} = \sqrt{H_{i}^{2}-K_{i}}\;.	
\end{equation}
Finally, energy minimization is performed via the Malitsky-Mishchenko adaptive gradient descent method~\cite{Malistky2020}, where the configuration $\bm{R}=\{\bm{r}_{1},\bm{r}_{2},\ldots\,\bm{r}_{N}\}$ of the triangular network is evolved by means of the following iteration rule:
\begin{equation}
\bm{R}(t+1) = \bm{R}(t)-l(t)\nabla F(t)\;,
\end{equation}
where $t$ is a time-like iteration counter, $\nabla=\{\nabla_{\bm{r}_{1}},\nabla_{\bm{r}_{2}},\ldots\,\nabla_{\bm{r}_{N}}\}$ and the step size $l(t)$ is chosen at each iteration as
\begin{align}
l(t) = \min\bigg\{
&l(t-1)\sqrt{1+\frac{l(t-1)}{l(t-2)}},\notag\\
&\frac{|\bm{R}(t)-\bm{R}(t-1)|}{2|\nabla F(t)-\nabla F(t-1)|}
\bigg\}\;.
\end{align}
The simulations are initialized in either one of the three configurations displayed in Fig.~\ref{fig:5}b-d and a displacement of magnitude $10^{-2}\ell_{0}$ and uniformly distributed random direction is added to the position of each vertex to compute $\bm{r}_{i}(0)$. We then take $l(0)=10^{-6}$ and $l(1)=0.5|\bm{R}(1)-\bm{R}(0)|/|\nabla F(1)-\nabla F(0)|$ to perform the first iteration. Our simulations continue until the averaged net displacement $\Delta(t) = l(t)|\nabla F(t)|/N$ is below $10^{-9}$.

\bibliography{tubulin_2022_09_26.bbl}

\end{document}